\documentclass[12pt]{article} \textheight=22 true cm
\usepackage{amssymb,amsmath,latexsym}
\usepackage{subfigure}
\usepackage{cite}
\usepackage{graphicx}
\DeclareMathOperator{\sech}{sech}
\usepackage{color}
\usepackage[colorlinks]{hyperref}
\begin{document}
\begin{center}
{\Large\bf Tolman-Bondi-Lema\^itre spacetime  with  a generalised
Chaplygin gas }\\[15 mm]
D. Panigrahi\footnote{ Sree Chaitanya College, Habra 743268, India
\emph{and also} Relativity and Cosmology Research Centre, Jadavpur
University, Kolkata - 700032, India , e-mail:
dibyendupanigrahi@yahoo.co.in }
  and S. Chatterjee\footnote{ Associate Professor(Retd.), New Alipore College, Kolkata - 700053, India \emph{and also}
Relativity and Cosmology Research Centre, Jadavpur University,
Kolkata - 700032, India, e-mail : chat\_sujit1@yahoo.com}\\[10mm]

\end{center}

\begin{abstract}
The Tolman-Bondi-Lema\^itre type  of inhomogeneous spacetime with generalised Chaplygin gas equation of state given by $p = -\frac{A}{\rho^{\alpha}}$ is investigated where $ \alpha$ is a constant. We get an inhomogeneous spacetime at early stage but  at the late stage of universe the inhomogeneity disappear with suitable radial co-ordinate transformation. For the large scale factor our model behaves like $\Lambda$CDM type which is in accord with the recent WMAP studies. We have calculated $\frac{\partial \rho}{\partial r}$ and it is found to be negative for $\alpha > 0$ which is in agreement with the observational analysis. A striking difference with Chaplygin gas ($ \alpha = 1$) lies in the fact that with any suitable co-ordinate transformation our metric cannot be reduced to the Einstein-de Sitter type of homogeneous spacetime in dust distribution as is possible for the Chaplygin gas.  We have also studied the effective deceleration parameter and find that the desired feature of \emph{flip} occurs at the late universe. It is seen that the flip time depends  explicitly on $\alpha$. We also find that flip is not synchronous occurring earlier  at the outer shells, thus offering a natural path against occurrence of wellknown shell crossing singularity.  This is unlike the Tolman-Bondi case with perfect gas where one has to impose stringent external conditions to avoid this type of singularity. We further observe that if we adopt  separation   of variables method to solve the field equations the inhomogeneity in matter distribution disappears.   The whole situation is later discussed with the help of Raychaudhury equation and the results  compared with previous cases. This work is the generalisation of our previous article where we have taken $\alpha =1$.

\end{abstract}

\bigskip
\section{ Introduction}
\vspace{0.1 cm}
From the growing number of observational data of high-redshift and
luminosity-distance relation of type IA supernovae  in the
last decade ~\cite{rei, aman}, we know that when interpreted in the framework of
Einstein's field equations and the standard FRW type of universe
(homogeneous and isotropic), we are left with the only alternative
that the universe is currently passing through an accelerated
phase of expansion where baryonic matter paradoxically contributes
only four percent of the total energy budget. Moreover if we have
faith in Einstein's theory the FRW model dictates that one should
hypothesize at once a peculiar and rather unphysical type of
matter field(DE)~\cite{kama} with a very large negative pressure clearly
violating the energy conditions, to explain the late acceleration. \\

 In the existing literature, a fairly a good number of DE models
are proposed, but very little is precisely suggested about the
nature and origin of it. Nowadays, the DE problem remains one of
the major open problems of theoretical physics ~\cite{wen}. On the way of
searching for possible solutions of this problem various models
are explored during the last few decades, referring to new exotic
forms of matter, \emph{e.g.}, quintessence ~\cite{chiba, cald}, phantom ~\cite{phan, phan1}, holographic
models ~\cite{hol}, string theory landscape ~\cite{string, string1 }, Born-Infeld quantum
condensate ~\cite{quantum}, the modified gravity approaches ~\cite{star, star1},
inhomogeneous spacetime ~\cite{krasin, sc1}, various types of higher dimensional
theories etc (readers interested in more detail for a
comprehensive overview of existing theoretical models may refer to ~\cite{bamba, li1, yoo,dp}). The one which attracted huge attention is the  Chaplygin
gas(CG) inspired model ~\cite{kam, dp1,dp2, dp3 }, obeying an EoS, $ p = -\frac{A}{\rho}$ .
Although the model is very successful in explaining the SNe Ia
data, it shows that the CG fails to explain  the tests connected
with structure formation and observed strong oscillations of
matter power spectrum ~\cite{sand}. To overcome  the problems it is
generalized (GCG) ~\cite{ben} with the addition of an arbitrary constant as

\begin{equation}\label{eq:eq15}
p = -\frac{A}{\rho^{\alpha}}
\end{equation}
where both $A (A>0)$ and $\alpha$ are constants. Here $\alpha$ is constrained in the
range $0 < \alpha < 1$ in order to have an acoustic speed that is at most luminal for perturbation ~\cite{fabris} and also  for best fit with observations ~\cite{camp, paul, dpbc}. Another
bottleneck stems from the fact that the basic inferences from the
$\Lambda$CDM  and GCG are essentially the same and so one can not
chose between the two from experimental angel. One more  point of
concern is the fact that the accelerating phase coincides with the
period when the inhomogeneities in the matter distribution at
length scales $ < 10 $ Mpc become significant so that the universe
can no longer be approximated as homogeneous at these scales.
Moreover one may point out that homogeneity and isotropy of the
geometry are not essential ingredients to establish a number of
relevant results in relativistic cosmology. One need not be too
sacrosanct about these concepts so as to sacrifice basic physics
(energy conditions, for example) in relativistic cosmology. On the
other hand if we forgo the concepts of homogeneity and isotropy,
the observational data do not force us to imply an accelerating
expansion of the universe, or even if the cosmic expansion is
accelerating it does not necessarily point to an existence of a
dark energy. Thus a parallel line of activities has emerged to
explain the observational findings without introducing the concept
of dark energy.  A community of activists have started a sort
of 'mission' to explain (sometimes with conflicting claims) the
observational findings within inhomogeneous models. Given the
complexities involved in dealing with inhomogeneous models the
simplest generalisation of FRW spacetime is the wellknown Tolman-Bondi-Lema\^itre
model which is also spherically symmetric but the spacetime is
inhomogeneous and the acceleration is  supposedly caused by the  back reaction
effects due to the inhomogeneities in the background FRW universe.
 It was shown that from observational point of view their~\cite{krasin, dp}
results become very similar to the predictions of CDM model.\\

The  motivation for the present work may be summed up as follows: As pointed out earlier that following the discovery of the late acceleration of our cosmos and the subsequent inability of the standard models to explain the phenomenon within the context of Einstein`s theory with standard perfect fluid there have been a proliferation of proposals to reintroduce the idea of a cosmological constant, a quintessential field, higher dimensional theories, higher derivative models etc. But all these suffer from the disqualification that the exotic fluid violate energy conditions and  also not physically viable.\\

An alternative line of approach is to address the problem in the realm of inhomogeneous cosmology, such that the back reaction coming from the extra terms due to inhomogeneity may trigger and drive the acceleration without being forced to invoke the presence of any exotic fluid and a vigorous search for compatibility of late acceleration with inhomogeneous model ensued~\cite{kolb}. But the journey is not free from controversies and failures. Returning  to the idea of back reaction Kolb \emph{et al.}~\cite{kolb} argued, using perturbative techniques, that when observed from the centre of perturbation the expansion rate is large and sometimes may accelerate. The work later got credence   from similar analysis of Wiltshire~\cite{wil} and also Carter \emph{et al.}~\cite{cart} where the universe is modeled as underdense bubble in an Einstein- de Sitter universe and predictions tally with those of $\Lambda$CDM. However it is later pointed out~\cite{giov, alnes} that the claim is seriously flawed as  domain of validity of perturbation is extrapolated to a regime where perturbative analysis breaks down as also constraints are violated.\\

So it points to the fact that acceleration can not be explained with the help of inhomogeneities alone. Therefore we have thought it fit to explore the phenomenon of late acceleration in inhomogeneous model with the help of now popular Chaplygin gas to see if the two are compatible i.e. if one can explain acceleration in this framework also.

As is common in all Chaplygin types of models our field equations
are amenable to closed form solutions only at the extremal cases.
Unlike the FRW models  all the physical parameters are here both
space and time dependent and all our solutions reduce to our
earlier work ~\cite{dp2} when $\alpha = 1$.\\

The organization of work is as follows : in section $ 2 $  we write
the field equations of our inhomogeneous spacetime with a
generalized Chaplygin gas as matter field and find the detail
solutions in section $3$. The solution described by our equation \eqref{eq:eq38}
 is unique and may be termed as generalised Einstein-de Sitter
metric(ED) and one can not directly revert to the well known ED
metric with any coordinate transformation. At the late stage of evolution
we get the solution similar to $\Lambda$CDM model. We also calculate the
acceleration flip in our spacetime, which depends both on space
and time. Evidently flip is not synchronous like homogeneous case.
Each shell characterised by a $r$-constant hypersurface has its own
instant of flip.\\

For any inhomogeneous dynamics we come across two
important singularities - shell crossing and shell focussing. We
have noted that in our case shells with higher value of $r$ starts
accelerating earlier and so shell crossing singularity is
naturally avoided. For completeness we contrast our inferences with
those obtained from Raychawdhury equation ~\cite{ray} and the paper ends with
a brief discussion in section $4$.

\section{ Field equations and its integrals}

\begin{equation} \label{eq:eq1}
  ds^{2}= dt^{2}- e^{\lambda (t,r)}~dr^{2}-R^{2}(t,r) \left( d\theta^{2}+\sin^{2}\theta d\phi^{2} \right)
\end{equation}
where the scale factor, $R(t,r)$  depends on both time and space
coordinates $(t,r)$ respectively. As inhomogeneous equations in GTR are, in general, very difficult to solve analytically we assume for mathematical simplicity that $g_{00}=1$.

 \vspace{0.1 cm}
 In comoving coordinate  system the energy momentum tensor for the above
defined coordinates is given by

\begin{equation}\label{eq:eq3}
T^{\mu}_{\nu} = (\rho + p)\delta_{0}^{\mu}\delta_{\nu}^{0} -
p\delta_{\nu}^{\mu}
\end{equation}
where $\rho(t,r)$ is the matter density  and $p(t,r)$ is the
pressure. The fluid consists of successive shells marked by $r$,
whose local density  is time-dependent over the successive
hypersurfaces. The function $R(t,r)$ describes the location of the
shells characterized  by $r$ at the time $t$. Einstein's field
equations, subject to the rescaled gauge
\begin{equation}\label{eq:eq4}
 R(0, r) = r
 \end{equation}

gives the following independent  equations for the metric
\eqref{eq:eq1} and the energy momentum tensor \eqref{eq:eq3} as

\begin{equation}\label{eq:eq5}
- \frac{e^{-\lambda}}{R^2} \left (2RR'' +R'^2 - RR' \lambda' \right) + \frac{1}{R^2} \left(R\dot{R} \dot{\lambda} + \dot{R}^2 + 1 \right)= \rho
\end{equation}

\begin{equation}\label{eq:eq6}
   -e^{- \lambda}\frac{R'^2}{R^2} +  \frac{1}{R^2}\left(2R\ddot{R} + \dot{R}^2 +1  \right)                                                 = -p
\end{equation}

\begin{equation}\label{eq:eq7}
 \frac{e^{-\lambda}}{R^2} \left(2RR'' +R'^2 - RR' \lambda' \right) + \frac{1}{R^2}
  \left(R \dot{R} \dot{\lambda} +\dot{R^2}+1 \right) = -p
\end{equation}

\begin{equation}\label{eq:eq8}
2 \dot{R'} - \dot{\lambda} R'= 0
\end{equation}
Here prime and a dot overhead denotes space and time derivative
respectively.

Solving equation \eqref{eq:eq8} we get

\begin{equation}\label{eq:eq9}
 e^{\frac{\lambda(t,r)}{2}} =\frac{R'}{f(r)}
\end{equation}

where $f(r)$ is an arbitrary function of $r$ such that $ f(r)
>0$.

 \vspace{0.2 cm}
Since the WMAP and other recent data \cite{rasa, hin} point to a nearly
flat universe in the current era we take $f(r)=1$ such that the
field equations finally reduce to the following two independent
equations as

\begin{equation}\label{eq:eq10}
\frac{\dot{R}^{2}}{R^{2}}+ 2 \frac{\dot{R'}}{R'}\frac{\dot{R}}{R}=\rho
\end{equation}

\begin{equation}\label{eq:eq11}
2\frac{\ddot{R}}{R}+ \frac{\dot{R}^{2}}{R^{2}}= -p
\end{equation}

The conservation equation leads to

\begin{equation}\label{eq:eq14}
\frac{d\rho}{dt} +
\frac{1}{e^{\frac{\lambda}{2}}R^{2}}\frac{d}{dt}\left( e^{\frac{\lambda}{2}}R^{2}\right)(\rho +p)=0
\end{equation}

For our case we take a matter field,  given by equation \eqref{eq:eq15} along with \eqref{eq:eq14} we get

\begin{equation}\label{eq:eq16}
\dot{\rho} +
\frac{1}{e^{\frac{\lambda}{2}}R^{2}}\frac{d}{dt}\left(e^{\frac{\lambda}{2}}R^{2}\right)\left(\rho -
\frac{A}{\rho^{\alpha}}\right) = 0
\end{equation}

which, on integration, gives

\begin{equation}\label{eq:eq17}
\rho = \left[ A + \frac{C(r)}{(e^{\frac{\lambda}{2}}R^{2})^{1+\alpha}}\right]^{\frac{1}{1+\alpha}}
\end{equation}

where $C(r)$ is a function of integration. Now putting equation \eqref{eq:eq9}, we get

\begin{equation}\label{eq:eq18}
\rho = \left[ A + \frac{C(r)}{(R' R^{2})^{1+\alpha}}\right]^{\frac{1}{1+\alpha}}
\end{equation}

With the help of equation  \eqref{eq:eq10} we finally get

\begin{equation}\label{eq:eq19}
\frac{\dot{R}^{2}}{R^{2}}+ 2 \frac{\dot{R'}}{R'}\frac{\dot{R}}{R}
= \left[ A + \frac{C(r)}{(R' R^{2})^{1+\alpha}}\right]^{\frac{1}{1+\alpha}}
\end{equation}
This is the main equation  in our future analysis but unlike the
homogeneous models, $C(r)$ depends on space also. As is well known
the resulting field equations with Chaplygin type of matter field
do not, in general,  offer any closed type of solutions and in what follows we
see that we have to study some extremal cases only. Following
Moffat ~\cite{mof1} the present authors, in an earlier communication ~\cite{dp2},
  have taken the expression of Hubble parameter as

\begin{equation}\label{eq:eq20}
H = \frac{2}{3} H_{\perp} + \frac{1}{3} H_{r}
\end{equation}
where

\begin{equation}\label{eq:eq21}
H_{\perp} = \frac{\dot{R}}{R}
\end{equation}
and
\begin{equation}\label{eq:eq22}
H_{r} = \frac{\dot{R'}}{R'}
\end{equation}
which may be taken as a measure of the local expansion rate in the
perpendicular and radial directions respectively. Now we can write
the deceleration parameter

\begin{equation}\label{eq:eq23}
q_{\perp} = - \frac{1}{H_{\perp}^{2}}\frac{\ddot{R}}{\dot{R}}
\end{equation}
From equation \eqref{eq:eq18} another important physical quantity,  $\frac{\partial
\rho}{\partial r}$ ( a sort of measure of inhomogeneity) comes out
to be

\begin{equation}\label{eq:eq24}
\rho' = \frac{\partial \rho}{\partial r} = - \frac{C(r)}{1+ \alpha} \frac{(1+\alpha) \left(\frac{R''}{R'} + 2 \frac{R'}{R} \right) - \frac{C'}{C(r)}}{\left(R^2 R' \right)^{1+\alpha} \rho^{\alpha}}
\end{equation}

For realistic mass distribution $\rho' <0 $ implying

\begin{equation}\label{eq:eq25}
(1+\alpha) \left(\frac{R''}{R'} + 2 \frac{R'}{R}\right) > \frac{C'}{C(r)}
\end{equation}
If we consider $C(r)$ to be a true constant then from equation
\eqref{eq:eq24}, we see that $\rho' <0$ as expected.
Otherwise we have to know the form of $C(r)$ to get an idea regarding the negativity of $\rho'$. We have chosen here two simple forms  of $C(r)$ as (i) power law \& (ii) exponential to check the negativity of $\rho'$ in the next section.\\
 \vspace{0.1 cm}

\section{Solutions}

As pointed out earlier the parent equation \eqref{eq:eq19} admit of hypergeometric solutions only in general. So we have to take some special cases only.

\textbf{CASE~A:}($R(t,r)$ is very small)

  When the scale factor $R(t,r)$ is relatively small, \emph{i.e.}, at the early stage of the universe, from equation \eqref{eq:eq19} we get dust dominated universe for $C(r) =
\left(\frac{4}{3} \alpha r^{3 \alpha -1} \right)^{1+\alpha}$
yielding

\begin{equation}\label{eq:eq36}
R(t,r)= r^{\alpha } \left[ t+ t_{0}(r)\right]^{\frac{2}{3}}
\end{equation}
where $t_{0}(r)$ is an arbitrary function of integration depending on $r$.

With this expression of $R (t,r)$ the pressure vanishes.  Moreover,
for isotropic expansion ($e^{\frac{\lambda}{2}}= R$) we get
$\rho\sim\frac{1}{R^{3}}$  (in an $r$-constant hypersurface) as in FRW universe. Interestingly the expression \eqref{eq:eq36} is not exactly
Tolman-Bondi like because we are dealing with a generalised Chaplygin gas type exotic fluid  and our line element reduces to

\begin{equation}\label{eq:eq37}
ds^{2}= dt^{2} - r^{2(\alpha -1)}\left[ t +
t_{0}(r)\right]^{\frac{4}{3}}\left \{\alpha^2 dr^{2} + r^{2}\left( d\theta^{2}+\sin^{2}\theta d\phi^{2} \right)\right \}
\end{equation}
If,  we further assume that $t_{0}(r)$ vanishes or  becomes a \emph{true} constant (in that case
a time translation is necessary) then we get

\begin{equation}\label{eq:eq38}
ds^{2}= dt^{2} - r^{2(\alpha-1)} t^{\frac{4}{3}} \left \{\alpha^2 dr^{2} + r^{2}\left( d\theta^{2}+\sin^{2}\theta d\phi^{2} \right)\right \}
\end{equation}
 The spacetime described by the equation \eqref{eq:eq38} is
 unique and one may look upon it as a modified Einstein-deSitter metric
 for the inhomogeneous spacetime. There is a striking difference between the spacetime described by equation \eqref{eq:eq38} and that in our work ~\cite{dp2} referred to earlier for $\alpha = 1$. In our previous work with pure Chaplygin gas ($\alpha = 1$) the additional assumption of $t_{0}(r) = 0$ reduces the metric to a homogeneous Einstein-de Sitter case with dust distribution in the flat space ($R^{\frac{2}{3}}$). But here $t_{0}(r) = 0$ does not reduce the metric to any homogeneous form. For that we need an additional assumption of $\alpha = 1$. So generalised Chaplygin gas does not admit of any homogeneous distribution in Tolman-Bondi metric.  From
equation \eqref{eq:eq18} we get the expression of density as

\begin{equation}\label{eq:eq39}
\rho(t,r)\approx \frac{\sqrt C(r)}{\left(R'R^{2}\right)^{1+\alpha}}= \frac{ 4 \alpha}{3r \left[t+
t_{0}(r)\right]\left[\alpha \frac{\{t + t_{0}(r)\}}{r}+
\frac{2}{3}t'_{0}\right]}
\end{equation}
If we calculate the deceleration parameter $q_{\perp}$ using
equations \eqref{eq:eq23} and \eqref{eq:eq36} we get $q_{\perp} =
\frac{1}{2}$ implying a dust dominated universe. From equation
\eqref{eq:eq39} we  have checked the signature of $\rho'$  given
by

\begin{equation}\label{eq:eq39a}
\rho' = -\frac{8 \alpha  \left[(3 \alpha +1) \left\{t_0(r)+t\right\} t_0'(r)+r t_0'(r){}^2+r \left\{t_0(r)+t\right\} t_0''(r)\right]}{\left\{t_0(r)+t\right\}{}^2 \left\{3 \alpha  t_0(r)+2 r t_0'(r)+3 \alpha  t\right\}{}^2}
\end{equation}
The equation \eqref{eq:eq39a} shows that $\rho'$ is always
negative for positive value of $\alpha$ as desired. This equation further
ensures that $\alpha $ should be greater than zero.

 \vspace{0.1 cm}
 \textbf{CASE~B :} ($R(t,r)$ is very large)

\textbf{Type - 1:}
 In the late stage of evolution the second term of the RHS of the
 equation \eqref{eq:eq19} vanishes and we get

\begin{equation}\label{eq:eq40}
\frac{\dot{R}^{2}}{R^{2}}+ 2 \frac{\dot{R'}}{R'}\frac{\dot{R}}{R}
=  A^{\frac{1}{1+\alpha}}
\end{equation}

(a) A straightforward integration of the  equation \eqref{eq:eq40} gives
$R(t,r)$ as

\begin{equation}\label{eq:eq41}
R (t,r) =  R_{0} \exp\left[\sqrt{\frac{A^{\frac{1}{1+\alpha}}}{3}} (t+r)\right]
\end{equation}
This is the wellknown de Sitter type of solution generalised to
inhomogeneous spacetime with  $A^{\frac{1}{1+\alpha}}$ simulating  as
$\Lambda$, the cosmological constant. It may be pointed out at this
  stage that the beauty of the idea of Chaplygin   gas lies in the
  fact that it unifies both the dark matter and dark energy concept
   in different limits producing an early dust dominated and an accelerating
   phase at the late stage of the evolution. It is found that this late stage
   expansion mimics the $\Lambda$CDM  model. At this stage a comparison
    to an earlier work of Moffat ~\cite{mof11} of LTB  model with cosmological constant may be relevant.
    Our key equation
   \eqref{eq:eq19} yields the solution \eqref{eq:eq41} for large scale factor which
    is strikingly similar to the Moffat result ~\cite{mof11}. However, the essential
     difference lies in the fact that while Moffat assumed \textit{apriori}
      a cosmological constant in his analysis but in our case
      it manifests itself at a late stage of evolution.
 Moreover a simple  radial
coordinate transformation

\begin{equation}\label{eq:eq42}
\bar{r}= R_{0}\exp\left[\sqrt{\frac{A^{\frac{1}{1+\alpha}}}{3}} r\right]
\end{equation}
reduces the metric \eqref{eq:eq1} to

\begin{equation}\label{eq:eq43}
ds^{2}= dt^{2} - \exp
\left({2~\sqrt\frac{A^{\frac{1}{1+\alpha}}}{3}~t}\right)\left\{ d\bar{r}^{2} +
\bar{r}^{2}
~\left( d\theta^{2}+\sin^{2}\theta d\phi^{2} \right) \right\}
\end{equation}
At the late stage of evolution it is seen that with suitable transformation of radial co-ordinate (equation   \eqref{eq:eq42} ) we get de-Sitter type metric with homogeneous spacetime. So it may be concluded that for large $R(t,r)$ the inhomogeneity may disappear as expected.

One can also see  from equation \eqref{eq:eq18} that for the late
universe

\begin{equation}\label{eq:eq44}
\rho \simeq   A^{\frac{1}{1+\alpha}}  + \frac{C(r)}{(1+\alpha)A^{\frac{\alpha}{1+\alpha}}}\frac{1}{(R'R^{2})^{1+\alpha}}
\end{equation}

\begin{equation}\label{eq:eq45}
p \simeq  - A^{\frac{1}{1+\alpha}}  + \frac{\alpha}{1+\alpha} \frac{C(r)}{A^{\frac{\alpha}{1+\alpha}}} \frac{1}{(R'R^{2})^{1+\alpha}}
\end{equation}
This may be viewed as a combination of a cosmological constant
$A^{\frac{1}{1+\alpha}}$ with a type of matter representing  a $\Lambda$CDM  model.  Moreover in the asymptotic limit($R
 \sim\infty$), we get $p = - \rho = - A^{\frac{1}{1+\alpha}}$ for this Chaplygin type of gas, corresponding to an empty universe with a cosmological constant.

  In this case the deceleration parameter $q_{\perp} = -1$, which shows an acceleration at the late stage. Now we can calculate $\rho'$ using equation \eqref{eq:eq44}, we get

\begin{equation}\label{eq:eq46}
\frac{\partial \rho}{\partial r} = - \frac{C(r)}{(1+\alpha) A^{\frac{\alpha}{1+\alpha}}} \frac{1}{(R'R^{2})^{1+\alpha}} \left\{ (1+\alpha) \left( \frac{R''}{R'} + 2 \frac{R'}{R} \right) - \frac{C'}{C(r)} \right\}
\end{equation}
 which is consistent with the inequality condition  \eqref{eq:eq25} for $\rho' < 0$. Now with the help of equation  \eqref{eq:eq41}, the condition \eqref{eq:eq25}  reduces to $\sqrt{3} (1+\alpha) A^{\frac{1}{2(1+\alpha)}} > \frac{C'}{C(r)} $. Since $C(r)$ is a positive integration constant, it may be true constant or may be a function of $r$. If the integration constant $C(r) \equiv C$ is a true constant then  $\rho' < 0$. On the other hand, if $C(r)$ depends on  $r$ such that $ C(r) \propto e^{\gamma r}$, which gives $\sqrt{3} (1+\alpha) A^{\frac{1}{2(1+\alpha)}} > \gamma $ and under this condition $\rho' <0$.\\

(b) Alternatively, one may  also get another type of solution of \eqref{eq:eq40} as

\begin{equation}\label{eq:eq46a}
R(t,r) = R_0 ~\sinh^{\frac{2}{3}} w(t+r)
\end{equation}
where $w = \frac{\sqrt{3}}{2} A^{\frac{1}{2(1+\alpha)}}$, unlike the previous work ~\cite{dp2} this result does not contain any explicit reference of $\alpha$, being absorbed in the expression of $w$.

\begin{figure}[ht]
\begin{center}
  \includegraphics[width=10cm]{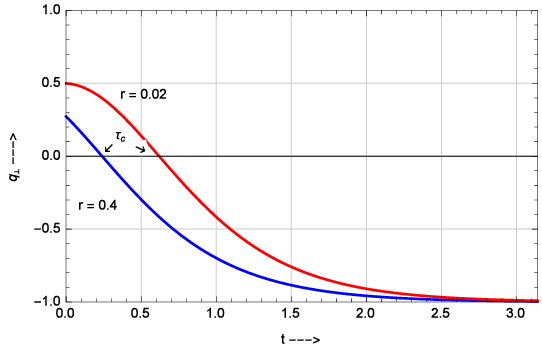}
  \caption{
  \small\emph{The variation of $q_{\perp}$ vs $t$  is shown in this
  figure. Taking $A=2$ \& $\alpha = 1$.  }\label{1}
    }
\end{center}

\end{figure}

Now using equations \eqref{eq:eq23} \& \eqref{eq:eq46a} we get the deceleration parameter as

\begin{equation}\label{eq:eq46b}
q_{\perp} = \frac{3}{2}  \sech^{2} w(t+r) -1
\end{equation}
Figure-1 shows that  the \emph{flip}  occurs early at greater value of $r$, \textit{i.e.},
 velocity increases for greater $r$. The flip time $\tau_{c}$ can be calculated
 from equation \eqref{eq:eq46b} when $q_{\perp} = 0$ and we get

\begin{equation}\label{eq:eq46c}
\tau_{c} = \frac{2}{\sqrt{3}} A^{- \frac{1}{2(1+\alpha)}} \text{sech}^{-1} \left(\sqrt{\frac{2}{3}}\right) -r
\end{equation}
As expected the flip time ($\tau_c$) explicitly depends on $\alpha$. The variation of $\tau_c$ with $\alpha$ depends
 on magnitude of $A$. If $A >1$, the $\tau_c$ increases as $\alpha$ increases, \emph{i.e.}, late flip for large
 $\alpha$,  on the other hand, for $A <1$,
 \emph{i.e.}, the conclusion is just the reverse.  For $A = 1$, ~ $\tau_c $  is independent on $\alpha$ for $r$-constant hypersurface.  The variation of
  $\tau_c$ with $\alpha$ for different values of $A$ are shown in figure-2.

 Another important conclusion coming out of the equation  \eqref{eq:eq46c} has not escaped our notice. As is customary in any inhomogeneous evolutions, this equation  shows that  all physical quantities including instant of flip depend on both space and time co-ordinate. So each shell characterised by a $r$-constant hypersurface has its own  flip time. Moreover, we further observe that shells with higher values of $r$ start accelerating earlier than those with lower values of $r$. This is a good news because it avoids the wellknown shell crossing singularity associated with any inhomogeneous evolution. This is unlike the Tolman-Bondi case with perfect gas where one has to impose stringent external conditions to avoid this type of singularity.

  Now we have to check the signature of $\rho'$. Using the condition \eqref{eq:eq25} we may write $(1+\alpha) \left \{ \tanh w(t+r) + \coth w(t+r) \right \} > \gamma$ for $\rho' < 0$ where we have taken $C(r) = e^{\gamma r}$.

\begin{figure}[ht]
\begin{center}
  \includegraphics[width=10cm]{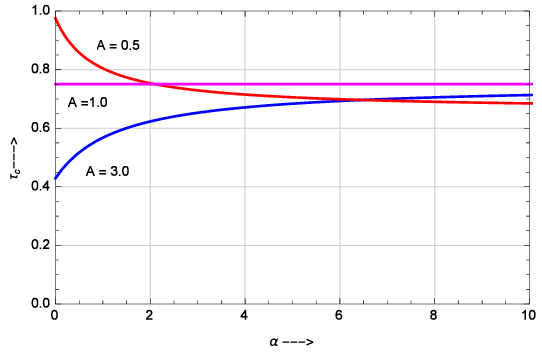}
  \caption{
  \small\emph{$\tau_c$ with $\alpha$  for different value od $A$ are shown in this
  figure. Taking $r=0.01$ .  }\label{1}
    }
\end{center}
\end{figure}
\vspace{0.5 cm}

\textbf{Type-2 :}
Now we attempt to solve the equation \eqref{eq:eq19} using the
method of separation of variables. Let $R(t,r) = a(t) g(r) $. From
equation \eqref{eq:eq19} we get

\begin{equation}\label{eq:eq47}
3 \frac{\dot{a}^2}{a^2} = \left( A +
\frac{B}{a^{3(1+\alpha)}}\right)^{\frac{1}{1+\alpha}}
\end{equation}
where

\begin{equation}\label{eq:eq48}
B = \frac{C(r)}{\left(g'g^2\right)^{1+\alpha}}
\end{equation}
But the LHS equation \eqref{eq:eq47} depends on time only which dictates that $B$ must be a
true constant.

 Now using equations \eqref{eq:eq24} \& \eqref{eq:eq48} a long but straight forward calculation
  shows that  $\rho' = 0$ ($C$ may be a function of $r$ or a true constant), implying that the matter field is homogeneous in this case.
 May not be out of space to point out that  one of the authors discussed, \emph{albeit} in a
 different context, the same situation and got similar results~\cite{sc2}.\\

\textbf{Temporal Solution :}\\

The equation \eqref{eq:eq47} gives the hypergeometric solution of $a(t)$ with $t$. The solution and other features are same as homogeneous case ~\cite{dp} at the late stage of evolution, i.e., $a(t)$ is large in this case, the equation \eqref{eq:eq47} becomes (neglecting higher order terms)

\begin{equation}\label{eq:eq49}
3 \frac{\dot{a}^2}{a^2} =   A^{\frac{1}{1+\alpha}} + \frac{B}{(1+\alpha) A^{\frac{\alpha}{1+\alpha}}}a^{-3(1+\alpha)}
 \end{equation}

\begin{figure}[ht]
\begin{center}
  \includegraphics[width=10cm]{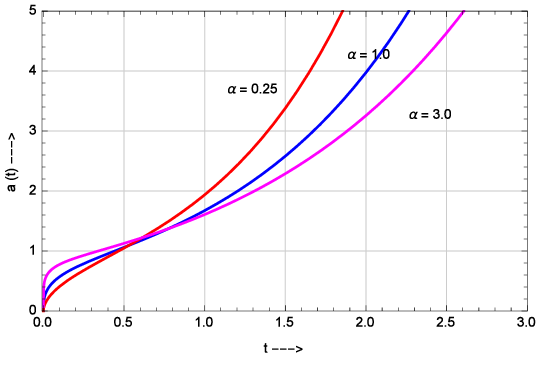}
  \caption{
  \small\emph{The variation of $a(t)$ vs $t$  is shown in this
  figure. Taking $A=5$ \& $B=5$.  }\label{1}
    }
\end{center}
\end{figure}
Solving the equation \eqref{eq:eq49} we get the solution,

\begin{equation}\label{eq:eq50}
a(t) = a_{0} \sinh^{m} \omega t
 \end{equation}

where, $a_{0} =  \left \{ \frac{B}{A(1+\alpha)} \right \}^{\frac{1}{3(1+\alpha)}}$ ; $m = \frac{2}{3(1+\alpha)}$ and $\omega = \frac{\sqrt{3}}{2} (1+\alpha) A^{\frac{1}{2(1+\alpha)}}$

\begin{figure}[ht]
\centering \subfigure[ Figure shows that the maximum value of $t_c$ at $\alpha = 0.2$. ]{
\includegraphics[width= 6.8 cm]{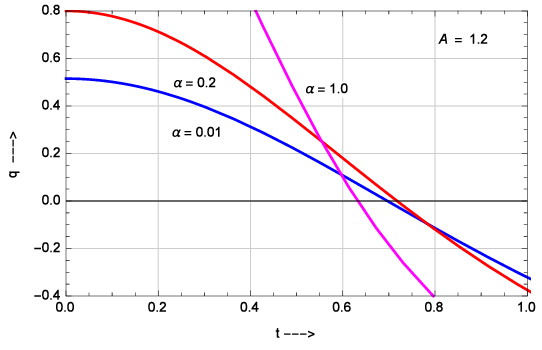}
\label{fig:subfig1} } ~~~\subfigure[ $t_c$ becomes maximum  at $\alpha = 0.255$.
 ]{
\includegraphics[width= 6.8 cm]{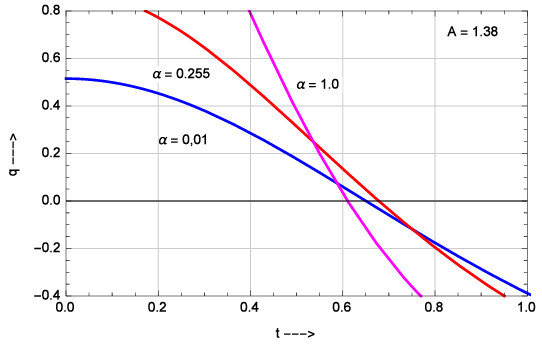}

 \label{fig:subfig2} }\label{fig:subfigureExample}~~~~~~~~~~~\caption[Optional
caption for list of figures]{\emph{ The variation of $q$ and $t$  for
different values of $\alpha$ with $B =1$.}}
\end{figure}

 From equation \eqref{eq:eq50}  we get the deceleration parameter

\begin{equation}\label{eq:eq52}
q = \frac{1-m \cosh^2 \omega t}{m \cosh^2 \omega t}
 \end{equation}
The equation \eqref{eq:eq52} shows that the exponent $m$ determines the evolution of $q$. A little analysis of equation \eqref{eq:eq52} shows that (i) if  $m > 1$ we get only  acceleration, no \emph{flip} occurs in this condition. But for $m > 1$ gives $- \frac{1}{3} > \alpha$, which is physically unrealistic, since previously we have shown $\alpha > 0$. (ii) Again, if $0< m< \frac{2}{3}$ it gives early deceleration and late acceleration and in this condition $\alpha > 0$, so the desirable feature of \emph{flip} occurs which agrees with the observational analysis for positive value  of $\alpha$.

Figure-4 shows the variation of $q$ with $t$ for different values of $\alpha$ where flip occurs. It is seen that the flip time ($t_c$) is different for different values of $\alpha$ but this change is not monotonous. We would like to focus on the occurrence of late flip as because all observational evidences suggest that accelerating phase is a recent phenomena. It is interesting to note that the late flip also depends on the value of $A$. In figure-4 we have taken two values of $A$ where we get the maximum $t_c$ for corresponding value of $\alpha$, \emph{e.g.}, for $A = 1.2$, we get the $(t_c)_{max}$ at $\alpha = 0.20$ and for $A = 1.38$, it comes out to be $\alpha = 0.255$. In this context correspondence to an earlier work of Campo ~\cite{camp} is relevant where he also got similar results while dealing with Generalised Chaplygin gas. Interesting to  mention that we also got similar results in our earlier work ~\cite{dpbc} although in a different context. From figure-4 we find that flip occurs later at this range of $\alpha$ in conformity with observational analysis.  Now The \emph{flip} time $(t_c)$ will be in this case

\begin{equation}\label{eq:eq53}
t_c = \frac{1}{\omega} \cosh^{-1} \left(\sqrt{\frac{1}{m}} \right)
 \end{equation}

 \begin{figure}[ht]
\begin{center}
  \includegraphics[width=10cm]{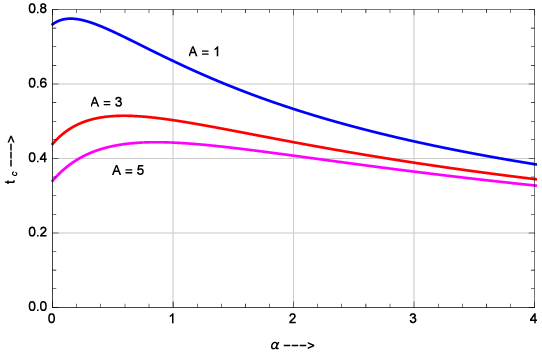}
  \caption{
  \small\emph{The graphs clearly show that flip time depends on  $\alpha$.   }\label{1}
    }
\end{center}
\end{figure}
Using equation \eqref{eq:eq53} we have drawn the figure-5 where the variation of $t_c$ with $\alpha$ for different value of $A$ is shown. It is seen that the variation of $t_{c}$ with $\alpha$ is not monotonous. When the value of $\alpha$ is small $t_{c}$ increases with $\alpha$; after a certain value of $\alpha$, $t_{c}$ decreases as $\alpha$ increases. That means we get a maximum value of $t_c$ for different value of $A$. As a trial case we see the following data table where we have seen the maximum value of $t_c$ for different value of $A$ with corresponding $\alpha$.

\vspace{0.6 cm}
\begin{table}[h!]
  \centering
  \caption{Table for  $\alpha$ and $t_c$}
  \label{tab:table1}
\begin{tabular}{|c|c|c|c|c|c|}
  \hline
  $A$ & 1.2 & 1.3 & 1.35 & 1.38 & 1.4 \\ \hline
  $(t_c)_{Max}$ & 0.7177 & 0.6945 & 0.6840 & 0.6780 & 0.6742 \\ \hline
  $\alpha$ & 0.200 & 0.233 & 0.246 & 0.255 & 0.260 \\ \hline
 \end{tabular}
\end{table}
From the above table we find that the value of $\left(t_c \right)_{max}$ is larger for smaller value of $A$.
From the observational point of view it is seen previously that this corresponds to a  value of $\alpha \sim 0.25$.  Table-1 further shows that for $\alpha = 0.255$,  the $\left(t_c \right)_{max}$ will be $ 0.6780$ when we consider the value of $A = 1.38$. \\

\vspace{0.1 cm}

\textbf{The radial solution:}

\vspace{0.1 cm}

  \begin{equation}\label{eq:eq54}
g(r) = \left\{ \frac{3}{B^{\frac{1}{1+\alpha}}} \int C(r)^{\frac{1}{1+\alpha}} dr \right\}^{\frac{1}{3}}
 \end{equation}

Now we have two options  - (i) $C(r)$ is a function of $r$ only \& (ii) $C(r) $ be a true constant : \\

(i) we may choose the simplest form of $C(r)$ :

 (a)$C(r) = r^{\beta}$, where $\beta $ is a constant. The equation \eqref{eq:eq54} reduces to

  \begin{equation}\label{eq:eq55}
  \int C(r)^{\frac{1}{1+\alpha}} dr = \frac{1 +\alpha}{1+ \alpha +\beta} r^{ \frac{1 + \alpha +\beta}{1 + \alpha}}
 \end{equation}
and we get

\begin{equation}\label{eq:eq55a}
  g(r) = \left\{\frac{3 (1+ \alpha)}{(1+ \alpha + \beta)B^{\frac{1}{\alpha+1}}}\right \}^{\frac{1}{3}} r^{\frac{ 1+ \alpha +\beta}{3(1+\alpha)}}
 \end{equation}

 (b) $C(r) = e^{\gamma r}$, where $\gamma$ is a constant.

  \begin{equation}\label{eq:eq56}
  \int C(r)^{\frac{1}{1+\alpha}} dr = \frac{1 + \alpha}{\gamma} e^{ \frac{\gamma r}{1 + \alpha}}
 \end{equation}
which gives

 \begin{equation}\label{eq:eq56a}
  g(r) = \left\{\frac{3( 1 + \alpha)}{\gamma B^{\frac{1}{\alpha+1}}}\right \}^{\frac{1}{3}} e^{ \frac{\gamma r}{3(1 + \alpha)}}
 \end{equation}

 (ii) When $C(r)$ is a true constant, i.e., $C(r) \equiv C$, the expression of $g(r)$ is give by

 \begin{equation}\label{eq:eq56b}
  g(r) = 3^{\frac{1}{3}} \left( \frac{C}{B} \right)^{\frac{1}{3(1+\alpha)}} r^{\frac{1}{3}}
 \end{equation}

 \vspace{0.1 cm}
\textbf{ The general solution :}
 \vspace{0.1 cm}

 Now the general solution will be

 \begin{equation}\label{eq:eq57}
 R(t,r) = \left[\frac{3}{\{A(1+\alpha)\}^{\frac{1}{1+\alpha}}} \int C(r)^{\frac{1}{1+\alpha}} dr \right]^{\frac{1}{3}}  \sinh^m \omega t
  \end{equation}
Using equations \eqref{eq:eq55},\eqref{eq:eq56} \& \eqref{eq:eq57} we can write the general solution in the following form

(i)$C(r)$ is a function of $r$ :\\

(a)$C(r) = r^{\beta}$ :\\

\begin{equation}\label{eq:eq58}
 R(t,r) = \left\{\frac{3 }{(1 + \alpha +\beta )}\right \}^{\frac{1}{3}} \left\{\frac{(1+\alpha)^{\alpha}}{A}\right\}^{\frac{1}{3(1+\alpha)}} r^{\frac{1+\alpha+  \beta}{3(1+\alpha)}} \sinh^m \omega t
  \end{equation}

  (b) $C(r) = e^{\gamma r}$ :\\

  \begin{equation}\label{eq:eq59}
 R(r,t) = \left\{\frac{3}{\gamma }\right \}^{\frac{1}{3}} \left\{\frac{(1+\alpha)^{\alpha}}{A}\right\}^{\frac{1}{3(1+\alpha)}} e^{ \frac{\gamma r}{3(1+\alpha)}}  \sinh^m \omega t
  \end{equation}
  and

  (ii) $C(r) \equiv C$ :\\

   \begin{equation}\label{eq:eq59a}
 R(r,t) = 3^{\frac{1}{3}} \left\{ \frac{C}{A(1+\alpha)} \right\}^{\frac{1}{3(1+\alpha)}} r^{\frac{1}{3}}  \sinh^m \omega t
  \end{equation}
when we put $\beta = 0$ in the equation \eqref{eq:eq58}, $C(r)$ becomes constant (unity) and equations \eqref{eq:eq58} \& \eqref{eq:eq59a} are identical.

  If we calculate  both $q_{\perp}$ and $t_{c}$, we get the same expressions  \eqref{eq:eq52} \& \eqref{eq:eq53} respectively because we are using the method of separation of variables to calculate the solution of $R(t,r)$.

It is to be mentioned that we have considered here  $C(r)$  is proportional to both power law and exponential function of $r$. Actually, these type of assumptions based on some solutions of $R(t,r)$,\emph{e.g}., in equation   \eqref{eq:eq36} we see that $R(r) \propto r^{\alpha}$, on the other hand, we get exponential relation in the equation  \eqref{eq:eq42}; in a different work of  Moffat ~\cite{mof2} got the same type of  exponential function of $r$.

\vspace{0.1 cm}

 \vspace{0.1 cm}
\section{ Raychaudhuri Equation }
\vspace{0.1 cm}

For sake of completeness we have contrasted the  results obtained so far with those obtained from the
 well known Raychaudhuri equation ~\cite{ray}, given by

\begin{equation}\label{eq:eq60}
\theta_{,\mu}v^{\mu} = \dot{v}^{\mu}_{; \mu} - 2(\sigma^{2}-
\omega^{2})-\frac{1}{3}~~ \theta^{2} + R_{\nu
\eta}v^{\nu}v^{\eta}
\end{equation}
where the terms have their usual significance. For our
irrotational system it reduces to

\begin{equation}\label{eq:eq61}
  \theta^{2}q = 6 \sigma^{2}+ 12 \pi G \left( \rho
  + 3p \right)
\end{equation}
With the help of the equations \eqref{eq:eq15}, \eqref{eq:eq18} \&
\eqref{eq:eq61} we finally get for deceleration parameter

\begin{equation}\label{eq:eq62}
  \theta^{2}q = 6 \sigma^{2}+ 12 \pi G \left[ -2A + \frac{C(r)}{(R'
  R^{2})^{1+\alpha}} \right]\left[ A + \frac{C(r)}{(R' R^{2})^{1+\alpha}} \right]^{-\frac{\alpha}{1+\alpha}}
\end{equation}
and for shear scalar

\begin{equation}\label{eq:eq63}
\sigma^{2} = \frac{1}{2}\sigma_{\mu\nu}\sigma^{\mu \nu} = \frac{1}
{3}\left(H_{r} - H_{\perp}\right)^{2}
\end{equation}

\textbf{CASE~A : Early Stage:} At the early phase of this
evolution when the scale factor $R(r,t)$ is small enough the
equation  \eqref{eq:eq62} reduces to

\begin{equation}\label{eq:eq64}
  \theta^{2}q = 6 \sigma^{2}+ 12 \pi G~ \frac{~~\left[C(r) \right]^{\frac{1}{1+\alpha}}}{R'R^{2}}
\end{equation}
It follows from the equation \eqref{eq:eq64} that $q$, the deceleration
factor is always positive. So  accelerated expansion is absent  in
this dust dominated phase though inhomogeneity is present here. The same conclusion was obtained previously   using equation \eqref{eq:eq36} where $q_{\perp} = \frac{1}{2}$.
Interestingly this result is very similar to  the work of Alnes
\emph{et al }~\cite{alnes}.

 \vspace{0.1 cm}
 \textbf{CASE~B : Late Stage :}

 \vspace{0.1 cm}
 \textbf{Type - I:} If we consider
the late stage of evolution \emph{i.e.}, $R(t,r)$ is large enough in this
phase, the second term of the RHS of the equation \eqref{eq:eq19} vanishes
and we get from  equation \eqref{eq:eq62},

\begin{equation}\label{eq:eq65}
  \theta^{2}q = 6 \sigma^{2}- 24 \pi G A^{\frac{1}{1+\alpha}}
\end{equation}

(a) When we use  the scale factor given by equation \eqref{eq:eq41} the shear scalar becomes $\sigma^2=0$. The equation \eqref{eq:eq65} reduces to

\begin{equation}\label{eq:eq66}
  \theta^{2}q = - 24 \pi G A^{\frac{1}{1+\alpha}}
\end{equation}

It gives  accelerating universe at the late stage. In the previous section we get the same conclusion  with the help of equation \eqref{eq:eq41} where the value of $q_{\perp} = -1$.

(b) Again, if we consider the expression of the scale factor is given by the equation \eqref{eq:eq46a} the shear scalar becomes $\sigma^{2} =
\frac{8}{3} \omega^{2} \mathrm{cosech} ^{2} \left[2 \omega \left(
r+t \right) \right]$ \& $A = (\frac{4}{3}\omega )^{(1+\alpha)}$. The equation
\eqref{eq:eq65} reduces to

\begin{equation}\label{66a}
  \theta^{2}q = 16 \omega^{2} \mathrm{cosech} ^{2} \left[2 \omega \left(
r+t \right) \right] - 32 \pi G \omega^{2}
\end{equation}

\begin{figure}[h]
\begin{center}
  \includegraphics[width=10cm]{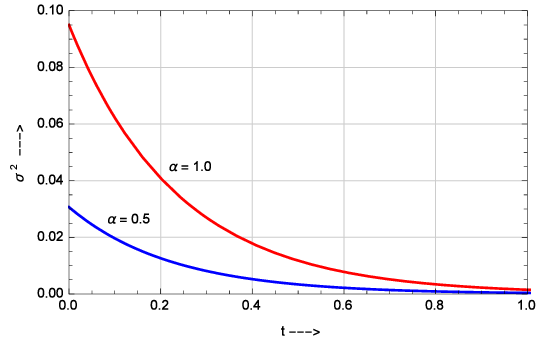}
  \caption{
  \small\emph{The variation of $\sigma^{2}$ vs $t$  is shown in this
  figure. Taking $A=2$ \& $r=1$.  }\label{1}
    }
\end{center}
\end{figure}

In figure-6  shows $\sigma^{2}$ vs $t$ for  $r$-constant hypersurface. In
this graph we have seen that as $t$ increases $\sigma^{2}$
decreases, \emph{i.e.}, when $t \rightarrow \infty $, $\sigma^{2}
\rightarrow 0$. So  initially it represents  the decelerating universe  and
after \emph{flip} we get acceleration  in line with current
observational result. It is to be mentioned that the expressions of $\sigma^2$ and $\theta^2 q$  seem to be identical  with our previous work ~\cite{dp2} but exactly not the same because here the expression of  $\omega$ contains $\alpha$.

\textbf{Type - II:} Again if we consider first order approximation
of equation \eqref{eq:eq57}, neglecting higher order terms, we get

\begin{equation}\label{eq:eq67}
  \theta^{2}q = 6 \sigma^{2}+ \frac{24 \pi G}{A^{\frac{\alpha}{1+\alpha}}} \left[ -A + \frac{1+3 \alpha}{2(1+\alpha)}\frac{C(r)}{(R'R^{2})^{1+\alpha}} \right]
\end{equation}
If we consider $R(t,r) = a(t)g(r)$, then from equation \eqref{eq:eq58} it follows that $\sigma = 0$.
 Now the equation \eqref{eq:eq62} reduces to

\begin{equation}\label{eq:eq68}
 \theta^{2}q =  \frac{24 \pi G}{A^{\frac{\alpha}{1+\alpha}}} \left[ -A + \frac{1+3 \alpha}{2(1+\alpha)}\frac{B}{a^{3(1+\alpha)}}\right]
\end{equation}
It follows from the equation \eqref{eq:eq68} that flip occurs when $a(t) =
\left\{\frac{1+3 \alpha}{2(1+\alpha)}\frac{B}{A} \right\}^{\frac{1}{3(1+\alpha)}}$. Now $q < 0$, at $a(t)
> \left\{\frac{1+3 \alpha}{2(1+\alpha)}\frac{B}{A} \right\}^{\frac{1}{3(1+\alpha)}}$ \emph{i.e.}, acceleration takes place in this case.

So we get early deceleration and late acceleration here. This also follows from equation \eqref{eq:eq52} for $\alpha >0$.\\

\section{Concluding Remarks}

We have considered a Tolman-Bondi-Lema\^itre type of inhomogeneous
spacetime with a generalised Chaplygin gas equation of state.
There is a proliferation of articles on accelerating universe with
Chaplygin EoS in homogeneous spacetime but scant attention has
been paid so far to address the problem in inhomogeneous
spacetime. But one intriguing problem is that accelerating phase
supposedly starts at the period when inhomogeneities in the
distribution in the universe at length scale $ < 10$ Mpc can no longer
be ignored. This primarily motivates us to investigate the matter
in inhomogeneous spacetime.  The salient features of other findings may be briefly summed up as:

(i) Our field equations being highly nonlinear with contributions
 from both inhomogeneity and generalised Chaplygin type of matter
field we have been able to get the solutions in closed form at
extreme cases only, \emph{i.e.}, at early and late stages of the
universe. In the former case we
 have seen that $\frac{\partial \rho}{\partial r}$ is always negative for $\alpha
 > 0$. From the theoretical point of view, we may conclude that the $\alpha$
 should be positive which is in agreement with the observational analysis. Here
 $C$ is a function of $r$, \emph{i.e.}, $C(r) = \left(\frac{4}{3} \alpha r^{3 \alpha
  -1}\right)^{1+\alpha}$. Interestingly we have seen that the deceleration
   parameter $q_{\perp} = \frac{1}{2}$ represents dust dominated universe.\\

(ii) In a different context the scale factor $R(t,r)$ has been
calculated at asymptotic range \emph{i.e.}, at late stage of the
universe. At the extreme case with suitable transformation of radial
co-ordinate the solution resembles de-Sitter type
 metric with homogeneous spacetime (see equation \eqref{eq:eq43}). So it may
 be concluded that at late stage of the universe inhomogeneity may disappear as  expected.

Further the integration function $C$ may be either a true constant
or a function of $r$. If we consider $C$ as a true constant then
$\frac{\partial \rho}{\partial r} < 0$ as desired for a regular
distribution in each case. Otherwise if $C \equiv C(r)$,
we have to take particular forms of $C(r)$ and $\rho' $ may be negative under certain restriction.\\

(iii) Another area of interest is the spacetime described by equation  \eqref{eq:eq38}. This is a unique result in the sense that for pure Chaplygin gas ($\alpha = 1$) one can reduce the equation  \eqref{eq:eq38} to the wellknown Einstein de-Sitter case with some additional assumption. However for the generalised Chaplygin gas ($\alpha \neq 0$) similar assumption does not reduce it to any homogeneous spacetime. \\

(iv) From equation \eqref{eq:eq46a} it further follows  that at the late era when \emph{flip} occurs, the flip time ($\tau_c$) depends explicitly on $\alpha$. The variation of $\tau_c $ with $\alpha$  also depends on magnitude of $A$ (figure-1). In this case flip occurs later for inner shells. \\

As is wellknown in an inhomogeneous model all physical
parameters depend on both space and time, including flip evidently the
time. It is not synchronous. The different shells characterised by
$r$-constant hypersurfaces start accelerating at different instants
of time. We have come across the phenomena of shell crossing
singularity in inhomogeneous gravitational collapse. But for an
inhomogeneous expanding model with acceleration this is
particularly significant. Because our analysis shows that for a
shell with a larger value of `$r$' the velocity flip starts
\emph{earlier}, a good news for avoidance of \emph{shell crossing
singularity}. So Chaplygin gas inspired model offers a natural
path against this singularity as opposed to the Tollman-Bondi case with
\emph{\emph{perfect gas}} where one has to impose a set of
stringent external conditions. \\

(v) For the sake of completeness, we have adopted the separation of variable method to solve our key equation  \eqref{eq:eq19}.    Most of the authors explained Chaplygin gas considering  extreme cases for temporal part. We have also studied the extremal form in Case A and Case B. Now for large $R(t,r)$ we consider upto second term of the temporal part and then we are able to solve the equation \eqref{eq:eq41} in exact form. The solution  of equation \eqref{eq:eq41} was shown in equation \eqref{eq:eq42} which shows early deceleration as well as late acceleration. The desirable feature of flip occurs which agrees with the observational analysis for positive value of $\alpha$.  In this case we find that the matter density becomes homogeneous \emph{i.e}., $\rho' = 0$ independent of the nature of $C$.\\

(a) One can also mention that the flip time ($t_c$) depends on the value of $\alpha$ but the dependance is not monotonic.
Figure-4 shows the variation of $q$ with $t$ for different values of $\alpha$ where flip occurs.
We have  concentrated on the occurrence of late flip as because all observational probes point to a late accelerating phase. It is interesting to mention that the late flip also depends on the value of $A$. In figure-4 we have taken two values of $A$ where we find the maximum $t_c$ for corresponding value of $\alpha$.

(b) To get the exact solution of the radial part represented by \eqref{eq:eq54} we have to choose the expression of integration constant $C(r)$ as the simplest form \emph{(i)}$C(r) = r^{\beta}$ \& \emph{(ii)} $C(r) = e^{\gamma r}$.  But if we consider $C(r)$ is a true constant, interestingly, we get $R(t,r) \propto r^{\frac{1}{3}}$, i.e., $R(t,r)$ is related to the power law expression of $r$.\\

(vii) We also have calculated $\theta ^2 q$ with the help of
Raychaudhury equation and showed that nature of $q$ is same for each case as in
section 3. \\

(viii) We further notice that in literature there exist models
generalising  LTB  with a cosmological constant. Our work essentially differs in that it is more general in nature because for a large scale factor, it reduces to that $\Lambda$CDM model where $A^{\frac{1}{1+\alpha}}$ simulates $\Lambda$ in equation \eqref{eq:eq41}.\\

As commented earlier in the introduction the Chaplygin Gas scenario,  besides its successful applicability in the accelerating universe paradigm, is also aesthetically satisfying in the sense that it beautifully synthesizes both matter and dark energy in a single whole unlike the $\Lambda$CDM case which explains only a part of the evolution. Moreover many workers including the present authors have also shown that the CG gas is thermodynamically stable ~\cite{dp4} as well.  But one should also point out that the CG cosmology also suffers from serious shortcomings in its attempts to explain the large scale structure formation of the universe, inviting serious comments and criticisms. Without going into details (those interested in more details are referred to references ~\cite{sand} and ~\cite{ben}) we would like to mention that the value of  the square sound velocity $c_s^2$  here comes out to be  very small which is shown to produce unphysical oscillations giving rise to finally an exponential growth of current power spectrum of matter ~\cite{sand}. However  recent analysis have shown that one can circumvent this difficulty taking the generalized Chaplygin Gas ~\cite{ben}. Moreover under the $\Lambda$CDM case the $c_s^2$  here though tiny but remains positive throughout.\\

The present work suffers from another serious short coming in that we have not so far attempted to constrain the  free model parameters with the help of observational data as is customary in relevant works in this field. The issue of compatibility of the obtained results with observational data will be addressed in our future work.

\textbf{Acknowledgments}

 \vspace{0.1 cm}
 DP acknowledges the financial support of Diamond Jubilee grant of Sree Chaitanya College, Habra.


\vspace{0.2 cm}



\begin{thebibliography}{50}


\bibitem{rei} A. G. Reiss et al,  \emph{Astron. J.} \textbf{607} 665
(2004).

\bibitem{aman} R. Amanullah \emph{et al},  \emph{Astrophy. J.} \textbf{716}
712 (2010).


\bibitem{kama} A. Kamada \emph{et al},  \emph{Phy. Rev. Lett.} \textbf{119}
111102 (2017).

\bibitem{wen} S. Weinberg, \emph{Rev. Mod. Phys.} \textbf{61},1(1989).

\bibitem{chiba} T. Chiba, N. Sugiyama and T. Nakamura, \emph{Mon. Not. Roy.
Astron. Soc.} \textbf{289} L5 (1997); [arxiv:astro-ph/9704199]

 \bibitem{cald} R.Caldwell, R. Dave and P. J. Steinhardt, \emph{Phys. Rev. Lett.}
\textbf{80}  1582 (1998); [arxiv:astro-ph/9708069].

\bibitem{phan} R. R. Caldwell, R. Dave and P. J. Steihardt \emph{Phys. Rev. Lett.}\textbf{80} 1582 (1998)
[arxiv:astro-ph/9908168].


\bibitem{phan1} R. R. Caldwell, \emph{Phys. Lett.}\textbf{B545} 23 (2002);
[arxiv:astro-ph/9908168].


\bibitem{hol} M. Li, \emph{Phys. Lett.} \textbf{B603} 1 (2004); [arxiv:hep-th/0403127].

\bibitem{string} L. McAllister and E. Silverstein, \emph{Gen. Rel. Grav.} \textbf{40} (2008) 565;  [arXiv:0710.2951]

\bibitem{string1} J. Polchinski, 2006 ;[arXiv:hep-th/0603249].


\bibitem{quantum} E. Elizalde, J. E. Lidsey, S. Nojiri and S. D. Odintsov, \emph{Phys. Lett.} \textbf{B 574}  1 (2003);
[arXiv:hep-th/0307177].

\bibitem{star} A. Shafieloo, V. Sahni and A.A. Starobinsky,
\emph{Phys. Rev. D} \textbf{80}  101301 (2009); [arXiv:0903.5141];


\bibitem{star1} M. I. Wanas,  Dark energy: Is it of torsion origin?, in
Proc. First \emph{MEARIM}, eds. A. A. Hady and M. I. Wanas(2009);
[arXiv: gr-qc/1006.0476v1]


\bibitem{krasin} A. Krasi´nski, C. Hellaby, M.-N. C´el´erier and K. Bolejko, \emph{Gen. Rel. Grav.} \textbf{42}
 2453 (2010) [arXiv:0903.4070]

\bibitem{sc1}  S. Chatterjee, \emph{JCAP} \textbf{03} 014 (2011); [arXiv:1012.1706]


\bibitem{bamba}  K. Bamba et al, \emph{Astrophysics Space Sc.} \textbf{342} 155 (2012); [arXiv:1205.3421]

\bibitem{li1}  M li, X D Li, S Wang and Y wang ,2012; [arXiv:1209.0922].

\bibitem{yoo}  J. Yoo and Y. Watanable,  \emph{Int. Jour. Mod. Phys.} \textbf{D 21} 1230002 (2012); [arXiv:1212.4726].

\bibitem{dp} D. Panigrahi and S. Chatterjee \emph{Gen Rel. Grav.} \textbf{40} 833 (2008); [arXiv:grqc /0709.0374];

\bibitem{dp1} D. Panigrahi and S. Chatterjee \emph{Grav. Cosml.} \textbf{17} 81
 (2011); [arXiv:grqc /1006.0476]; D Panigrahi , Proceedings No: CP  1316, Search
 for fundamental theory, edited by R L Amoroso, P Rowlands and S
 Jeffers , AIP, (2010)

\bibitem{kam}A. Kamenschik, U. Moschella and V. Pasquier, \emph{Phys. Lett.} \textbf{B511} 265 (2001)

\bibitem{dp2} D. Panigrahi and S. Chatterjee, \emph{JCAP} \textbf{10} 002(2011).

\bibitem{dp3} D. Panigrahi and S. Chatterjee, \emph{Int. Jour. Mod. Phys.} \textbf{D 21}, 1250079 (2012)

\bibitem{sand} H. Sandvik, M. Tegmark, M. Zaladarriaga and I. Waga ,  \emph{Phys.Rev. D}\textbf{69} 123524 (2004)


\bibitem{ben} M. C. Bento, O. Bertolami and A. A. Sen,  \emph{Phys.Lett.}\textbf{b 575} 172 (2003)

\bibitem{fabris} J. C. Fabris, S. V. B. Goncalves and De Souza E. V.  \emph{Gen. Rel. Grav.}\textbf{34} 53 (2002)

\bibitem{camp} S. del Campo,  \emph{JCAP} \textbf{11} 004(2013);[arxiv:astro-ph.CO/1310.4988]

\bibitem{paul} B. C. Paul and P. Thakur, \emph{JCAP} \textbf{11} 052(2013); [arxiv:astro-ph.CO/1306.4808]

\bibitem{dpbc} D. Panigrahi, B. C. Paul and S. Chatterjee, \emph{Grav. Cosmol.} \textbf{21}, 83 (2015);[ arxiv:gr-qc/1305.7204]


\bibitem{kolb} E. W. Kolb, S. Matarrese, A. Notari, A. Riotto, [hep-th/0503117]


\bibitem{wil} D. L. Wilshire, [gr-qc/0503099]

\bibitem{cart} B. M. N. Carter et al, [astr-ph/0504192]

\bibitem{giov} M. Giovannini, \emph{Phys. Lett.} \textbf{B634} 1 (2006)

\bibitem{alnes} H. Alnes, A. Morad  and \O~ Gron, \emph{JCAP} \textbf{01} 007 (2007)

\bibitem{ray} A. K. Raychaudhuri, Phys. Rev. \textbf{98} 1123 (1955)

\bibitem{rasa} S. Rasanen, JCAP \textbf{11} 003 (2006 );[arxiv:astro-ph /0607626]

\bibitem{hin} G. Hinshaw \emph{et al}, Astrophys. J. Suppl.  \textbf{208} 19 (2013) [arxiv:astro-ph.CO /1212.5226]

\bibitem{mof1} J. W. Moffat, JCAP \textbf{05} 001 (2006 ); [arxiv: astro-ph/ 0505326]

\bibitem{mof11} J. W. Moffat,  [arxiv: astro-ph/ 0504004]


\bibitem{sc2} S. Chatterjee and B. Bhui  \emph{Mon. Not. Roy. Astro. Soc.} \textbf{247} 57
 (1990)

\bibitem{mof2} J. W. Moffat (2006); [arxiv: astro-ph/0606124]

\bibitem{dp4} D. Panigrahi  \emph{Int. Jour. Mod. Phys.} \textbf{D 24} 1550030 (2015); [arXiv:grqc /1405.7667]



\end{thebibliography}
\end{document}